\documentclass[prb,twocolumn,superscriptaddress,amssymb]{revtex4-1}
\usepackage[utf8]{inputenc}
\usepackage[T1]{fontenc}
\usepackage{amsmath}
\usepackage{graphicx}
\usepackage{dcolumn}
\usepackage{bm}
\usepackage{epsfig}
\usepackage{float} 
\usepackage{color}
\usepackage{tikz}
\usepackage{subcaption}
\usepackage{verbatim}
\usepackage{float}
\usepackage[normalem]{ulem}
\usepackage{multirow}
\usepackage[colorlinks=true,urlcolor=blue,linkcolor=blue,
            citecolor=blue]{hyperref}

\newcommand \beq{\begin{equation}}
\newcommand \eeq{\end{equation}}

\newcommand \exciting{\texttt{Exciting}}

\newcommand \evec{{\bf e}}
\newcommand \kvec{{\bf k}}

\newcommand \qvec{{\bf q}}
\newcommand \pvec{{\bf p}}

\newcommand\rvec{{\bf r}}

\newcommand\Gvec{{\bf G}}

\newcommand{\bra}[1]{\langle #1|}
\newcommand{\ket}[1]{|#1\rangle}
\newcommand{\rhot}{\tilde{\rho}}

\newcommand{\alo}{$\alpha$-Al$_2$O$_3$}

\def\simge{\mathrel{%
       \rlap{\raise 0.511ex \hbox{$>$}}{\lower 0.511ex \hbox{$\sim$}}}}
\def\simle{\mathrel{
       \rlap{\raise 0.511ex \hbox{$<$}}{\lower 0.511ex \hbox{$\sim$}}}}
\def\beq {\begin{equation}}
\def\eeq {\end{equation}}

\def\w {\omega}
\def\bfq {\mathbf{q}}

\begin{document}

\title{Connections between resonant inelastic x-ray scattering and complementary x-ray spectroscopies: probing excitons at Al K and L$_1$ edges of \alo}

\newcommand{\lsi}{LSI, CNRS, CEA/DRF/IRAMIS, \'Ecole Polytechnique, Institut Polytechnique de Paris, F-91120 Palaiseau, France}
\newcommand{\etsf}{European Theoretical Spectroscopy Facility (ETSF)}
\newcommand{\soleil}{Synchrotron SOLEIL, L'Orme des Merisiers, Saint-Aubin, BP 48, F-91192 Gif-sur-Yvette, France}

\date{\today}

\author{M. Laura Urquiza}
\affiliation{\lsi}
\affiliation{\etsf}

\author{Matteo Gatti}
\affiliation{\lsi}
\affiliation{\etsf}
\affiliation{\soleil}

\author{Francesco Sottile}
\affiliation{\lsi}
\affiliation{\etsf}

\begin{abstract}

We present an {\it ab initio} study of neutral core and valence  electronic excitations  in {\alo} by solving the Bethe-Salpeter equation (BSE) of many-body perturbation theory within an all-electron framework. 
Calculated spectra at the Al K and L$_1$ edges are in remarkable agreement with available experiments from X-ray absorption (XAS) and X-ray Raman spectroscopy once excitonic effects are taken into account.
The combination of the BSE spectra for the two techniques confirms the dipole-forbidden nature of the exciton prepeak as suggested by recent calculations based on density-functional theory. 
Moreover, we make predictions for resonant inelastic X-ray scattering (RIXS)
spectra at K and L$_1$ edges, which strikingly fully overlap also beyond an independent-particle picture.

The RIXS calculations reveal two distinct regimes as a function of incoming photon energy. 
Below and at the XAS threshold, we observe Raman-like features, characterised by strong excitonic effects, which we directly compare to peaks in the loss function.
Above the XAS threshold, instead, fluorescence features become predominant: RIXS spectra can be well described and analyzed within an independent-particle approximation showing similarity with the X-ray emission spectrum.

\end{abstract}


\maketitle


\section{\label{sec:introduction} Introduction}

The past decades have witnessed huge progress  in both the experimental resolution~\cite{Zimmermann_2020,Ament_2011,deGroot_2001} and the computational accuracy \cite{Rehr_2000,Rehr_2006,deGroot_2021} 
of X-ray spectroscopies that probe core levels  in materials. Core-level spectroscopies\cite{DeGroot2008} have  thus become a key tool for the study of a vast number of materials properties, which is evidenced by the surge of interest in their applications across chemistry, physics, biology, and materials science~\cite{Lamberti_Bokhoven}. 

The fundamental processes of photon absorption, emission and scattering give rise to prominent spectroscopies that measure neutral electronic excitations in materials\cite{Huotari2012}.
X-ray absorption (XAS), also referred as X-ray absorption near-edge spectroscopy, is determined by electronic transitions from core levels to unoccupied states, while X-ray emission (XES) stems from the decay of valence electrons to fill a core hole, providing complementary information on occupied states. 
Finally, resonant inelastic X-ray scattering\cite{Schulke2007} (RIXS) results from the coherent combination of X-ray absorption and emission. In RIXS, the energy of an incoming   photon is tuned to resonate with a specific core level, and the subsequent relaxation of a valence electron to fill the core hole is accompanied by the emission of a photon of lower energy. As a result, RIXS probes low-energy excitations of the various degrees of freedom (i.e., charge, spin, lattice) of materials\cite{Ament_2011, Kotani_2001}.
If, instead, the incident photon energy is much higher than the typical binding energies of core levels, one has the non-resonant inelastic X-ray scattering\cite{Schulke2007} (NRIXS),  also called X-ray Raman scattering (XRS) when used to measure core-level properties.

These X-ray spectroscopies share  attractive properties.
The large penetration depth of photons, especially in the hard X-ray regime, allows better bulk sensitivity than for spectroscopies that make use of electrons, such as photoelectron spectroscopy.
Moreover, the use of the distinctive atomic transitions (i.e., the absorption edges) of the different kinds of atoms  ensures chemical sensitivity, providing element and orbital specific information of the local chemical environments in complex materials.
In particular, the selection rules of the photoexcitation make it convenient 
to identify the character of the electronic excitations.
Within an independent-particle picture, XES and XAS spectra can be simply related, respectively, to the  occupied and unoccupied projected density of states (PDOS) of the absorbing atom with the angular momentum component that fulfils the selection rules. 
Similarly, RIXS spectra can be associated to the projected joint density of states (JDOS).

However, electron-electron interactions can dramatically alter this independent-particle picture. In particular, according to the semiempirical final-state rule\cite{vonBarth_1982,Rehr_2005},  one has to deal with  the 
strong perturbing potential due
the presence of a core hole in the final state.
Indeed, the electron-hole attraction gives rise to core excitons in XAS and XRS, and to both core and valence excitons in RIXS. 
Excitons manifest as strong enhancements of the spectral weight at the onsets and are often even the main feature in the spectra.

In order to take into account the effects of the electronic interactions, 
a variety of methods have been developed at different levels of approximation, ranging from the real-space multiple-scattering formalism~\cite{XAS_Zabinsky_1995,NRIXS_Soininen_2005,RIXS_Kas_2011}, to
cluster models~\cite{XAS_Josefsson,XAS_Haverkort,XAS_Maganas} 
and many-body perturbation theory~\cite{RIXS_Shirley_1998,RIXS_Shirley_2000,RIXS_Shirley_2001,NRIXS_Vinson_2011,RIXS_Vinson_2016,RIXS_Vinson_2017,RIXS_Vinson_2019,RIXS_Vorwerk_2020,RIXS_Vorwerk_2022,RIXS_Gilmore_2021} (MBPT). Within the context of MBPT, the first-principles solution of  Bethe-Salpeter equation~\cite{Strinati_1988,Hanke_1979} (BSE) is nowadays the state-of-the-art approach to deal with excitonic effects for both core and valence excitations\cite{Bechstedt2014,Martin2016}. 
The BSE within an all-electron framework\cite{Exciting_2019} will be  therefore adopted also in the present study. 

The aims of the present work are to conduct an in-depth analysis of RIXS spectra, while accounting for coherence and excitonic effects throughout the entire process, and to establish direct connections with the complementary spectroscopy techniques that also measure neutral excitations in materials, such as XAS, XRS and XES. 
Our study is focused on \alo, which is a prototypical wide-band gap insulator with a broad range of applications, including catalysis, ceramics, and electronics\cite{Catalysis_2019,RRAM_2004}.

Since the BSE has consistently demonstrated its accuracy in describing similar experiments, our conclusions can be reliably extended to other wide band-gap materials.

We have calculated XAS and XRS spectra at the Al K edge reproducing very well the available experimental spectra\cite{K_XAS_Manuel_2012,K_XAS_Fulton_2015,K_NRIXS_Delhommaye} when strong excitonic effects, which determine the main peak in the spectra, are properly taken into account. 
Consistently with results from literature, we find that an excitonic prepeak is also present in the XRS spectrum at large momentum transfer. This prepeak is not visible in the calculated XAS spectrum because dipole forbidden, while it is enabled in the experiments by the coupling with atomic vibrations.
Moreover, we demonstrate that both XAS and RIXS spectra at the Al L$_1$ edge are the same as the corresponding spectra at the Al K edge, which suggests that L$_1$ edges could be a valuable alternative in the soft X-ray regime to K edges that require hard X-rays. 
The analysis of RIXS spectra shows that excitonic effects are mostly visible for excitation energies smaller than (or similar to) the absorption edges, and how for higher excitations energies the RIXS spectra tend to XES spectra where excitonic effects are less relevant.

The article is organized as follows. In Sec.~\ref{sec:method}, we briefly present the basic theoretical concepts for the calculation of the spectra,
together with a summary of the computational details. 
In Sec.~\ref{ssec:k_edge} we discuss the calculated XAS and XRS spectra at Al K edge. RIXS spectra for K and L$_1$ edges are then compared and analysed in Sec.~\ref{ssec:rixs_k_l1}.
Finally, in Sec.~\ref{sec:conclusions}, we summarize the main conclusions and give an outlook of the present work. 

\section{\label{sec:method} Theoretical framework and computational details}

\subsection{\label{ssec:method_XAS} The Bethe-Salpeter equation for excitation spectra}

The BSE is an in principle exact equation for the electron-hole correlation function, which is directly linked to the various neutral electronic excitation spectra\cite{Bechstedt2014,Martin2016}. 
Within the GW approximation~\cite{Hedin_1965} and using a statically screened Coulomb interaction $W$, the BSE can be expressed as an eigenvalue problem for the two-particle excitonic  Hamiltonian~\cite{Onida_2002}:
$\bar{H}_{\rm exc} \bar{A}_\lambda = \bar{A}_\lambda \bar{E}_\lambda$.

The matrix elements of the excitonic Hamiltonian can be written in the basis of 
electron-hole transitions $t$ between Bloch orbitals\footnote{For simplicity, here we consider explicitly only a spin-unpolarised situation.}  $n_1{\bf k}_1 \to n_2{\bf k}_2$, which in the {\it ab initio} framework are usually calculated within the Kohn-Sham scheme\cite{KohnSham_1965} of density-functional theory (DFT).

In this basis the matrix elements read:
\begin{equation}
 	\bra{t} \bar{H}_{\rm exc} \ket{t'} = E_{t} \delta_{tt'} + \bra{t} \bar{v}_c-W \ket{t'}.
	\label{eq:BSE_H} 
\end{equation}

Here $E_{t}$ is the independent-particle excitation energy 
calculated in the GW approximation.
$\bar{v}_c$ is the Coulomb interaction without its long-range, macroscopic, component: 
\begin{equation}
    \bar{v}_c(\qvec+\Gvec) = 
    \begin{cases} 
     4\pi / \lvert \qvec+\Gvec \rvert^2 \text{ for  }  \Gvec\neq 0 \\
     0 \text{ for } \Gvec=0
    \end{cases}
\end{equation} 
where $\Gvec$ is a reciprocal-lattice vector and $\qvec$ is a wave vector in the first Brillouin zone.
The modified Coulomb interaction $\bar v_c$ enters the excitonic Hamiltonian as an exchange electron-hole repulsion and is responsible for crystal local field effect~\cite{Wiser,Adler}.
The screened Coulomb interaction $W$ is calculated in the random-phase approximation (RPA). It plays the role of the direct electron-hole attraction and is responsible for excitonic effects.

In the Tamm-Dancoff approximation \cite{Taylor1954} (which will be assumed henceforth), 
one considers only resonant transitions between occupied states $n_1$ and empty states $n_2$.
For XAS spectra $n_1$ is a core level $\mu$, while for optical spectra $n_1$ is a valence band $v$.
In both cases, the absorption spectra, which are described by the imaginary part  of the macroscopic dielectric function $\textrm{Im}\epsilon_M(\omega)$ in the long wavelength limit $\bfq\to0$, are obtained in terms of the excitonic eigenvectors $\bar A_\lambda$ and eigenvalues $\bar E_\lambda$ as\cite{Onida_2002}:
\begin{equation}
   \textrm{Im}\epsilon_M(\omega) =
   \lim_{\qvec\to 0}\frac{8\pi^2}{\Omega q^2} 
   \sum_\lambda \left| \sum_{t}  \bar{A}_\lambda^{t}\tilde{\rho}_{t}(\qvec) \right|^2 \delta(\w- \bar{E}_\lambda),
\label{eq:XAS_BSE}
\end{equation}
where $\Omega$ is the volume of the system and the oscillator strengths $\tilde{\rho}_{t}(\qvec)$ are defined as:
\begin{equation}
	\rhot_{t}(\qvec) = 
	\bra{n_1\kvec-\qvec}  e^{-i\qvec\cdot\rvec} \ket{n_2\kvec}.
	\label{eq:rhotw}
\end{equation}

By setting the direct electron-hole interaction $W$  to 0 in the excitonic Hamiltonian \eqref{eq:BSE_H}, one finds the absorption spectra within RPA. 
Moreover, by switching off both electron-hole interactions  $\bar v_c$ and $W$, the independent-particle approximation (IPA)  is retrieved and the absorption spectrum becomes:
\begin{equation}
	\textrm{Im}\epsilon_M(\omega) =
	\lim_{\qvec\to 0}\frac{8\pi^2}{\Omega q^2} 
	\left| \sum_{t}  \tilde{\rho}_{t}(\qvec) \right|^2 \delta(\w - E_t).
	\label{eq:XAS_IPA}
\end{equation}
The direct comparison between Eq. \eqref{eq:XAS_BSE} and Eq. \eqref{eq:XAS_IPA} shows that the electron-hole interactions affect the spectra in two ways: by modifying the peak positions, through a change of excitation energies from the interband transition energies $E_{t}$ to the excitonic energies $\bar E_\lambda$, and by altering the peak intensities, through the mixing of the independent-particle transitions 
that are weighted by the excitonic coefficients $\bar A_\lambda$.

In XES spectra, the transitions $t$ take place between occupied valence states $v$ and empty core levels $\mu$.
In this case, there are no electron-hole interactions since both initial and final states contain one hole, but no excited electron.
Therefore, XES spectra are usually calculated in the independent-particle picture\cite{XES_Mortensen_1997}. In the dipole limit one has:
\begin{equation}
I^{\rm XES} (\omega)
\propto  \lim_{\qvec\to 0}\frac{8\pi^2}{\Omega q^2} \left| \sum_{t}  \tilde{\rho}_{t}(\qvec) \right|^2 \delta(\w- E_t).
\label{eq:XES_IPA}
\end{equation}
In this approximation, one neglects the effect of the core hole on valence states\cite{Aoki2019}.

By further assuming that the oscillator strengths in Eqs. \eqref{eq:XAS_IPA} and \eqref{eq:XES_IPA} fulfill the dipole selection rule, but are transition-independent constants, the spectra would be equivalent to the PDOS for empty states in XAS and the PDOS for occupied states for XES.

If, at variance with Eq. \eqref{eq:BSE_H}, the excitonic Hamiltonian $ H_{\rm exc}$ includes the full Coulomb interaction $v_c$:
\begin{equation}
 	\bra{t} H_{\rm exc} \ket{t'}
	= E_{t} \delta_{tt'} + \bra{t} v_c-W \ket{t'},
\label{eq:BSE_H2} 
\end{equation}
one obtains the loss function, i.e., the inverse macroscopic dielectric function\cite{NRIXS_Gatti_Sottile}
\begin{equation}
   -\textrm{Im}\epsilon_M^{-1}({\bf q},\omega) 
   =  \frac{8\pi^2}{\Omega q^2} 
      \sum_\lambda \left| \sum_{t}  A_\lambda^{t}\tilde{\rho}_{t}(\qvec) \right|^2 \delta(\w- E_\lambda)
\label{eq:EELS_BSE}
\end{equation}
in terms of the excitonic eigenvectors and eigenvalues: $H_{\rm exc}A_\lambda = E_\lambda A_\lambda$.
By comparing the two excitonic Hamiltonians Eq. \eqref{eq:BSE_H} and Eq. \eqref{eq:BSE_H2}, we understand that the long-range component ${\bf G}=0$ of the Coulomb interaction $v_c$ gives rise to the difference between the macroscopic dielectric function $\epsilon_M$ in the ${\bf q} \to 0$ limit, see Eq. \eqref{eq:XAS_BSE}, and its inverse $\epsilon_M^{-1}$, see Eq. \eqref{eq:EELS_BSE}.
When the electronic states are localised, such as for core levels, the long-range component  ${\bf G}=0$ of the Coulomb interaction $v_c$ becomes ineffective in the excitonic Hamiltonian \eqref{eq:EELS_BSE}, and therefore the two spectra \eqref{eq:XAS_BSE} and \eqref{eq:EELS_BSE} coincide\cite{Onida_2002,Sottile_2005,Mizuno_1967}.

NRIXS usually investigates electron-hole transitions $t$ from valence bands $v$ to conduction bands $c$, while XRS focuses on transitions from core levels $\mu$ to conduction bands $c$.
In both cases, the excitation spectra  are described by the dynamic structure factor\cite{Schulke2007}, which is proportional to the loss function \eqref{eq:EELS_BSE}. 
A big advantage of scattering spectroscopies with respect to absorption is the possibility to probe electronic excitations as a function of the momentum transfer $\bfq$, going well beyond the dipole limit $\bfq\to0$.

In the case of RIXS\footnote{Here we consider only the so called direct RIXS, in contrast to indirect RIXS, where the core hole potential induces secondary excitations\cite{Ament_2011}. Moreover, we focus on the electronic excitations only.}, the excitonic Hamiltonian \eqref{eq:BSE_H2} has to be solved twice.
In the first BSE, one deals with a core exciton Hamiltonian $H_{\rm exc}A_{\lambda_\mu} = E_{\lambda_\mu} A_{\lambda_\mu}$, where the electron-hole
transitions $t_\mu$ are from core levels $\mu$ to conduction bands $c$.
In the second BSE, one has a valence exciton Hamiltonian: $H_{\rm exc}A_{\lambda_o} = E_{\lambda_o} A_{\lambda_o}$, where the 
transitions $t_o$ are from valence bands $v$ to conduction bands $c$. Here we consider only vertical interband excitations at the same $\bf k$ point, i.e., we calculate RIXS spectra in the long-wavelength limit.

RIXS spectra can be then obtained from the two sets of excitonic eigenvalues and eigenvectors as\cite{RIXS_Vorwerk_2020,RIXS_Vorwerk_2022}:
\begin{equation}
 I^{\rm RIXS}(\omega_1,\w)\propto
	\sum_{\lambda_o} \left| \sum_{\lambda_\mu} \frac{ t^{(1)}_{\lambda_\mu} \, \, t^{(2)}_{\lambda_o,\lambda_\mu}}{\omega_{1} - E_{\lambda_\mu} + i\Gamma/2} \right|^2 
 \delta(\omega-E_{\lambda_o}) \ ,
\label{eq:RIXS_BSE}
\end{equation}
Here the energy loss $\omega$ is equal to the difference between the incident photon energy $\omega_1$ and the emitted photon energy $\omega_2$: $\omega=\omega_1-\omega_2$, $1/\Gamma$ is the lifetime of the core hole and  the oscillator strengths are 
\begin{align}
	t^{(1)}_{\lambda_\mu} =& \sum_{{\mu}c\kvec} A^{{\mu}c\kvec}_{\lambda_\mu} \bra{c\kvec}\evec_1\cdot\pvec\ket{\mu\kvec}
\label{eq:t1} \\
	t^{(2)}_{\lambda_o,\lambda_\mu} = &   \sum_{vc\kvec}  \sum_{\mu} A^{vc\kvec}_{\lambda_o} \bra{\mu\kvec}\evec_2^{*}\cdot\pvec\ket{v\kvec} \left[A^{{\mu}c\kvec}_{\lambda_\mu}\right]^*,
\label{eq:t2}
\end{align}
with $\pvec$ the momentum operator and $\evec_1$ and $\evec_2$ the light polarization unity vectors\footnote{
Differently from Eq. \eqref{eq:rhotw} that adopts a longitudinal gauge, in Eq. \eqref{eq:t1}-\eqref{eq:t2} the oscillator strengths are written in the transverse gauge. 
The two gauges are equivalent in the long-wavelength limit\cite{DelSole_1993}, where the polarization unity vectors $\evec$ and the wave vectors $\bfq$ play the same role defining the direction of the perturbation. Therefore, the oscillator strengths $t^{(1)}_{\lambda_\mu}$  in Eq. \eqref{eq:t1} are equivalent to those in the XAS spectra from Eq. \eqref{eq:XAS_BSE}. Similarly, the oscillator strengths $\bra{\mu\kvec}\evec_2^{*}\cdot\pvec\ket{v\kvec}$ in Eq. \eqref{eq:t2} correspond to those entering XES spectra in Eq. \eqref{eq:XES_IPA}.}.

While the possible peak positions in the scattering spectra from Eqs. \eqref{eq:EELS_BSE} and \eqref{eq:RIXS_BSE} are the same, their intensities can be very different. For example, in contrast to NRIXS,  RIXS spectra  even in the dipole limit can display dipole-forbidden excitations, such as $d$-$d$ transitions\cite{Ament_2011,Kotani_2001,Rueff_2010}, since they stem from a different two-step process.
In general, RIXS spectra strongly depend on the excitation energies $\omega_1$ in the denominators of Eq. \eqref{eq:RIXS_BSE}. For fixed excitation energy $\omega_1$, RIXS spectra  are usually plotted and analysed as a function of the energy loss $\omega$, (i.e., as $I^{\rm RIXS}(\omega_1,\w=\omega_1-\omega_2)$),  or of the emission energy $\omega_2$ (i.e., as $I^{\rm RIXS}(\omega_1,\omega_2)$).

In the IPA the RIXS spectrum \eqref{eq:RIXS_BSE} becomes:
\begin{multline}
    I^{\rm RIXS}(\omega_1,\w)\propto
	\sum_{cv\kvec} \left| \sum_{\mu} \frac{   \bra{c\kvec}\evec_1\cdot\pvec\ket{\mu\kvec} \bra{\mu\kvec}\evec_2^{*}\cdot\pvec\ket{v\kvec}}{\omega_{1} - (\varepsilon_{c\kvec}-\varepsilon_{\mu\kvec}) + i\Gamma/2} \right|^2 \\
	\times \delta(\omega-(\varepsilon_{c\kvec}-\varepsilon_{v\kvec})) \ ,
\label{eq:RIXS_IPA}
\end{multline}
which is given by a combination of vertical transitions (i.e., at the same $\kvec$ point) between core levels of energies $\varepsilon_{\mu\kvec}$, valence bands $\varepsilon_{v\kvec}$, and conduction bands $\varepsilon_{c\kvec}$. If the peak intensities are further assumed to be constant in Eq. \eqref{eq:RIXS_IPA}, the RIXS signal becomes proportional to the JDOS (projected on the angular component selected by the excitation-disexcitation process).
When electron-hole interactions are weak, one may therefore relate RIXS to band-structure properties\cite{Ma1994,RIXS_Ma_1994,RIXS_Carlisle_1995,RIXS_Carlisle_1999,RIXS_Carlise_2000,RIXS_Shirley_2000,Kotani_2001}. On the contrary, excitonic effects also in RIXS spectra mix  the various interband transitions, blurring the picture based on the single-particle band structure.

\subsection{\label{ssec:method_details} Computational details}

The computational strategy of the present work follows the one that we have already successfully employed for {\alo} in Ref. \cite{L1_XAS_Urquiza}.
We have adopted the experimental lattice parameter~\cite{Newnham_1962} $a_{0} = 5.128$~\r{A} and angle $\alpha = 55.287$ in the rhombohedral primitive cell.
Calculations have been performed using the full-potential all-electron (AE) approach, as implemented in the \exciting~code~\cite{Exciting_2014,Exciting_2019}. 
The Kohn-Sham ground-state wave functions have been calculated within the local density approximation~\cite{KohnSham_1965} (LDA) of DFT. 
The Brillouin zone has been sampled with a $6\times6\times6$ $\kvec$-grid, using plane waves (PW) expansion with a cutoff energy of 12~Hartree. 
The AE approach includes muffin-tin (MT) spheres with radii of 2~bohr and 1.45~bohr for aluminum and oxygen, respectively. 

BSE calculations are performed on a $8\times8\times8$ $\kvec$-grid shifted  by (0.05, 0.15, 0.25). 
In XAS and RIXS simulations, BSE matrix elements  
are calculated with a cutoff $|\Gvec+\qvec|_{\rm max} = 4~a^{-1}_0$, maintaining a PW cutoff of 7~Hartree for the wave functions. 
XRS calculations are performed with a PW cutoff of 10~Hartree and $|\Gvec+\qvec|_{\rm max} = 7~a^{-1}_0$.

To obtain the RPA screening of $W$ we used the same parameters as in the BSE, including 100 conduction bands. 
The BSE Hamiltonian was constructed considering 12 (4) occupied states and 20 (60) unoccupied states for the optical (L$_1$ and K edge) spectrum.  
In order to take into account GW corrections\cite{Marinopoulos_2011}, the LDA band gap has been opened by a rigid scissor correction of 2.64~eV. 
Excitations spectra at the Al L$_1$ and K edges have been aligned with the experimental spectra, applying a downshift of the corresponding LDA core level energies of 17.18~eV and 62.9 eV, respectively.
For the XRS spectrum, the downshift is 61.2 eV.
XAS and XRS spectra for the Al K edge have been convoluted with a Lorentzian broadening of 0.7~eV to match the experimental broadening, bigger than the inverse core-hole lifetime of 0.42~eV~\cite{Fuggle_Inglesfield, Campbell}.

The K and L$_1$ RIXS spectra have been calculated with the BRIXS code~\cite{RIXS_Vorwerk_2020, Vorwerk_thesis}, considering the first (lowest-energy) 17000 and 80000 BSE eigenvectors and eigenvalues for the core and valence excitations, respectively, which gives a converged RIXS spectrum for an energy loss window of 20~eV. 

For a better comparison between RIXS spectra at Al K and L$_1$ edges we have used the same core-hole inverse lifetime $\Gamma$ of 0.2 eV. 
The RIXS and XES spectra have been plotted using a broadening of 0.1 eV.

\section{\label{sec:results} Results and discussion}

{\alo} is made of alternate layers of Al and O atoms that are stacked along the $z$ axis of our cartesian reference frame. The nature of the chemical bond is largely ionic: the valence bands have mostly O $2p$ character, while the bottom conduction band is mainly due to Al $3s$.
The wide band gap\cite{Will_1992,French_1994} of $\sim$9.6 eV concurs with the low dielectric constant\cite{Harman1994,Schubert2000} of $\sim$ 3 to give rise to strong excitonic effects for both valence and core excitations\cite{Marinopoulos_2011,L1_XAS_Urquiza}.

Since the final state of XAS is the intermediate state of RIXS, its analysis is propaedeutic  to understand RIXS spectra. 
Therefore, in Sec. \ref{ssec:k_edge} we will first analyse XAS and XRS spectra, before moving to the study of RIXS spectra in Sec. \ref{ssec:rixs_k_l1}.

\subsection{\label{ssec:k_edge} XAS and XRS at Al K edge}

\begin{figure}[h]
	\includegraphics[width=\columnwidth]{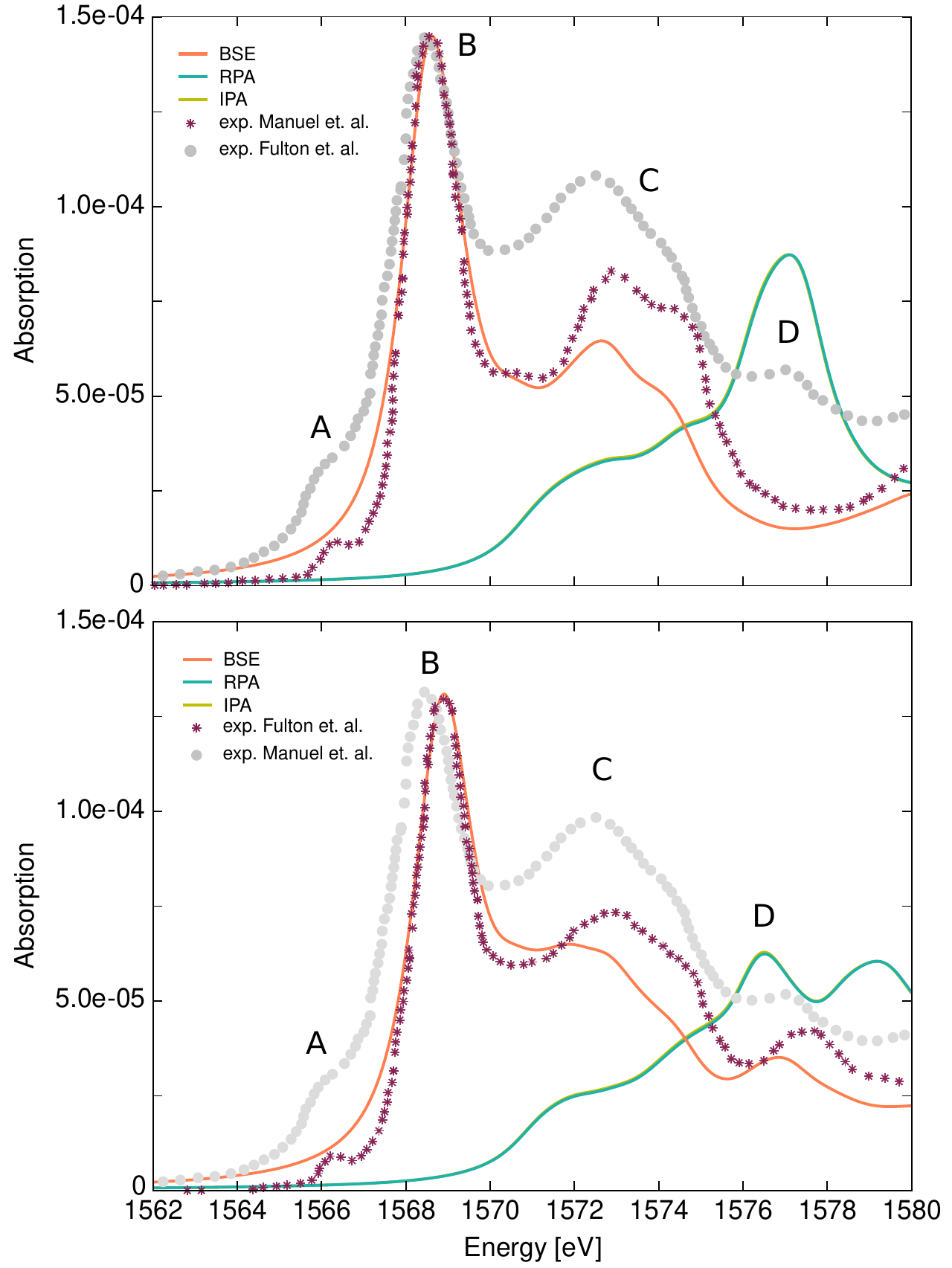}
	\caption[\columnwidth]{\label{fig:k_abs} Al K edge  absorption spectra calculated within the IPA, RPA and BSE for two polarization directions: (top panel) parallel ($xy$) and (bottom panel) perpendicular ($z$) to the {\alo} layers. 
	They are compared to two  experimental XAS spectra. While the spectra of Manuel {\it et al.}\cite{K_XAS_Manuel_2012} are polarization resolved, the spectrum of Fulton {\it et al.} \cite{K_XAS_Fulton_2015} represents an average of different polarization directions. 
	The experimental spectra have been  normalized to match the maximum of the BSE intensity. They have been further aligned at the main peak: the spectrum from Manuel~\textit{et al.}\cite{K_XAS_Manuel_2012} has been blueshifted by 1 eV.}
\end{figure}

Fig.~\ref{fig:k_abs} compares two experimental Al K edge XAS \cite{K_XAS_Manuel_2012,K_XAS_Fulton_2015} with the absorption spectra calculated within the IPA, the RPA and the BSE. The top and bottom panels display the spectra for two polarization directions: parallel ($xy$) and perpendicular ($z$) to the {Al$_2$O$_3$} layers, respectively.
While the spectrum of Fulton {\it et al.} \cite{K_XAS_Fulton_2015} represents an average of the polarization directions, the polarized XAS experiment by Manuel {\it et al.} \cite{K_XAS_Manuel_2012} has been done on a single crystal:  the two measured spectra distinguish the two orientations.

Like for the optical spectra and the shallower core edges\cite{L1_XAS_Urquiza},
the two Al K edge XAS spectra\cite{K_XAS_Manuel_2012}  display only a small anisotropy, which is very well captured by the BSE calculations. 
The IPA and RPA curves are on top of each other, meaning that the contribution of crystal local fields is negligible. This result may seem  surprising since Al $1s$ electrons are localised, giving rise to an inhomogeneous charge response that could be responsible for strong local fields\cite{Onida_2002}. However, $1s$ electrons are also not highly polarizable, which implies that the induced charge is so small that local field effects are negligible. 
On the contrary, taking into account excitonic effects in the BSE spectra is crucial to reproduce the main features of the experiments, notably the most prominent peak (B) and the secondary peaks (C) and (D).
The ratio between the features (B) and (C) is an important indicator for aluminium coordination\cite{K_XAS_Kato_2001,K_XAS_Mogi_2004}, used in particular to distinguish the tetrahedral AlO$_4$ from the octahedral AlO$_6$ of {\alo}.

In the experimental spectra there is also a prepeak (A) 
at $\sim 1566$~eV that is not present in the calculated spectra. In the BSE calculation, we actually find an excitonic eigenvalue $\bar E_\lambda$ at the same energy. However, its oscillator strength is negligibly small, i.e., it is a \textit{dark} exciton. Indeed, it corresponds to electron-hole transitions from the $1s$ core level to the bottom of the conduction band around the $\Gamma$ point with mostly $3s$ character, which therefore are dipole forbidden.
This prepeak (A) has been extensively studied  both theoretically and experimentally~\cite{K_XAS_LiPan1995,K_XAS_Cabaret_1996,K_XAS_Ildefonse1998,Mo_2000,K_XAS_Cabaret_2005,K_XAS_Cabaret2009,K_XAS_Brouder_2010,K_XAS_Manuel_2012,K_XAS_Nemausat_2016}. The current interpretation is that  atomic vibrations enhance the Al $sp$ hybridization at the bottom of the conduction band, by deviating the Al atoms from their centrosymmetric positions and thus enabling Al $1s\rightarrow3p$ atomic-like transitions in the measured XAS spectra.
Between the prepeak (A) and the main peak (B) we also identify additional excitons that have low oscillator strengths but are not completely dark. They correspond to transitions from $1s$ states to the first conduction band, for $\kvec$ points between $\Gamma$ and T, characterised by non-zero Al $sp$ hybridisation. 
  
The binding energy of the lowest energy dark exciton is 1.69 eV, while the three main exciton peaks (B), (C) and (D) are within the continuum of electron-hole transitions of the IPA spectrum.

\begin{figure}
	\includegraphics[width=\columnwidth]{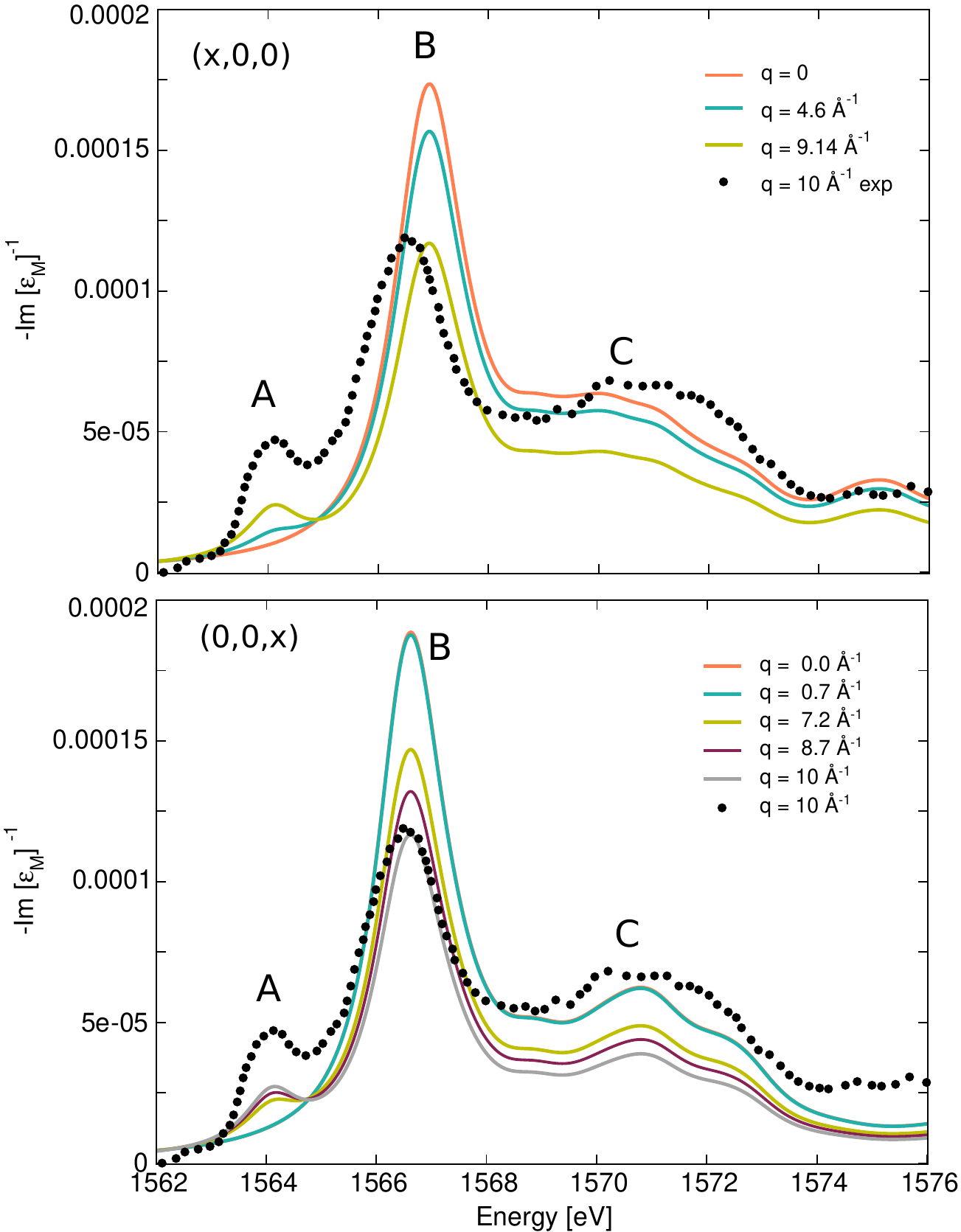}
	\caption[\columnwidth]{\label{fig:k_nrixs} Loss function calculated within the BSE for momentum transfers $\bfq$  along the $x$ (top) and $z$ (bottom) Cartesian directions, with a broadening of 0.7~eV.
 	The experimental XRS spectrum from Ref.~\cite{K_NRIXS_Delhommaye} was measured  for momentum transfer $ \vert\qvec\vert = 10$~\r{A}$^{-1}$.  
	Here, it has been scaled to match the maximum  intensity of the calculated spectrum at $ \vert\qvec\vert = 10$~\r{A}$^{-1}$. 
 	We note that the peak positions of the XRS spectrum\cite{K_NRIXS_Delhommaye} have a redshift of $\sim$1.7~eV with respect to the XAS spectra in Fig. \ref{fig:k_abs}.} 
\end{figure} 

In order to corroborate the interpretation that the prepeak (A) in XAS is associated to a dipole-forbidden exciton,  we also calculate the XRS spectra from the solution of the BSE at the same Al K edge.

In the dipole limit $\qvec\rightarrow0$, for localised electrons such as core levels,
the loss function (measured by XRS, see Eq. \eqref{eq:EELS_BSE}) tends to the absorption spectrum (measured by XAS, see Eq. \eqref{eq:XAS_BSE}):
\begin{equation}
	-\textrm{Im}\epsilon_M^{-1}(\omega)
	=  \frac{\textrm{Im}\epsilon_M(\omega)} {[\textrm{Re}\epsilon_M(\omega)]^{2}+ [\textrm{Im}\epsilon_M(\omega)]^{2}} \to \textrm{Im}\epsilon_M(\omega).
	\label{eq:nrixs-xas}
\end{equation}
Mathematically, this results from  $\textrm{Im} \epsilon_M(\omega) \ll \textrm{Re} \epsilon_M(\omega) \rightarrow 1 $. 
In this situation, the absorption spectrum and the loss function yield the same information.
As the magnitude of the momentum transfer $\qvec$ increases, instead, XRS  probes non-dipolar transitions, potentially leading to the emergence of features not visible in the XAS spectra. 

The BSE loss functions calculated for different momentum transfers $\qvec$  along the $x$ and $z$  Cartesian directions are shown in Fig~\ref{fig:k_nrixs}, and compared to the experimental XRS spectrum\cite{K_NRIXS_Delhommaye} for $q = 10$~\r{A}$^{-1}$. The measurement was done on a powder sample, therefore its momentum direction dependence could not be resolved. 
Also in this case the agreement between the BSE spectrum and experiment is noteworthy, and  could be further improved by taking into account the coupling with atomic vibrations\cite{K_NRIXS_Delhommaye}.

\begin{figure*}
	\includegraphics[width=1.5\columnwidth]{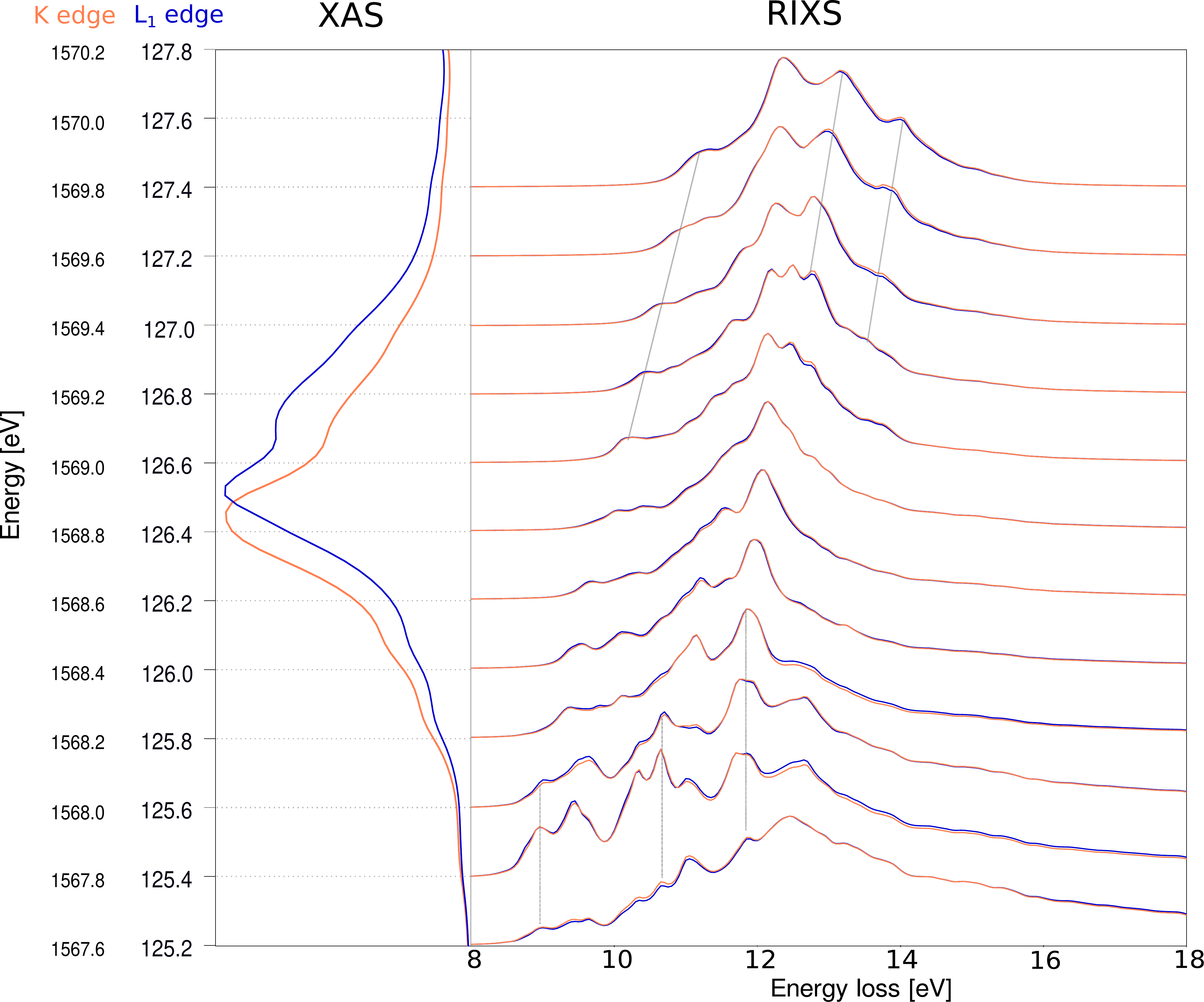}
	\caption[\columnwidth]{\label{fig:k_l1_ddcs} (Left) XAS spectra calculated in the BSE for the Al K and L$_1$ edges  (red and blue curves respectively) for polarization along the $xy$ direction. Note the different energy scales for the two edges. The spectrum for the K edge has been multiplied by 35.
	(Right) RIXS spectra at the K and L$_1$ edges (red and blue curves respectively) as a function of the energy loss $\w$, evaluated for excitation energies $\w_1$ (separated by 0.2~eV steps) across the corresponding XAS edge in the left panel. 
	The RIXS spectra have been calculated within the BSE for photon polarizations vectors $\evec_1$ and $\evec_2$ along the $x$ direction. 
	Each RIXS spectrum has been normalized to its maximum.
	The thin black lines connecting the different RIXS spectra are guides for the eye to identify Raman losses and fluorescence features.}
\end{figure*} 

In the dipole limit $\bfq\to0$, the loss functions match the corresponding absorption spectra in Fig. \ref{fig:k_abs}, as expected from the previous discussion. 
The loss function remains the same, without any trace of the prepeak (A), also for $q = 0.12$~\r{A}$^{-1}$, which corresponds to the momentum carried by X-ray photons absorbed at 1.5 KeV energy in the Al K edge XAS.
Instead, as the $q$ value further increases, the prepeak (A) becomes progressively more visible and intensifies in the loss function.
It becomes clearly detectable in the spectrum for $q \gtrsim 4$~\AA$^{-1}$. This behavior demonstrates unequivocally its non-dipolar origin.

Besides changing their relative intensities, the three main features (A), (B), (C) do not show appreciable dispersions as a function of the momentum transfer $\bfq$. Only the exciton peak (B) slightly disperses (by 0.1~eV) to higher energies for $\bfq$ along $z$.
The non-dispersive character of these excitations is a manifestation of their localised nature.

In summary, our BSE calculations highlight the strong core excitonic effects at the Al K edge of XAS and XRS.
The combination of absorption and loss function spectra strongly supports 
the non-dipolar Al $1s\to3s$ character of the exciton prepeak  (A). It becomes visible in the spectra either for a coupling with atomic vibrations or at finite momentum transfers $\bfq$.
Our results based on the solution of the many-body BSE agree with the single-particle DFT-based approach\cite{K_NRIXS_Delhommaye,K_XAS_Manuel_2012,K_XAS_Brouder_2010,K_XAS_Grad_2022}, where the final state is calculated for a supercell with a core-hole localised on the absorbing atom\footnote{Or, also, by using the $Z+1$ scheme\cite{K_XAS_Nakanishi_2009}.}.
The strong electron-hole attraction is explicitly accounted for in the BSE by the screened Coulomb interaction term $W$, and implicitly in the DFT approach by allowing for electronic relaxation in presence of a core hole\cite{Liang2017}.

\subsection{\label{ssec:rixs_k_l1} Resonant inelastic x-ray scattering at Al L$_1$ and K edges}

While, in principle, unoccupied $p$ states could be equivalently probed  in XAS   by excitations from either $1s$ or $2s$ core levels, theoretical and experimental investigations  predominantly focus on K edges disregarding L$_1$ edges. 
In the independent-particle picture,  excitation spectra of {\alo} at Al K and  L$_1$ edges can be described by  the same unoccupied Al $3p$ PDOS\cite{Nufer2001}.
When excitonic effects are taken into account, instead, one could expect that the higher degree of localisation of the $1s$ core hole could lead to stronger electron-hole interactions at the K edge than at the L$_1$ edge. Their impact on the spectra would be therefore different.

Instead, in the left panel of Fig. \ref{fig:k_l1_ddcs}, the comparison of the Al K and L$_1$ XAS spectra calculated within the BSE shows that their equivalence holds also beyond the independent-particle picture.
Indeed, the two calculated spectra are extremely similar, besides a small shift ($<$ 0.1 eV) of the main peak, and an overall scaling factor of $35$ of the intensity of the L$_1$ edge with respect to the K edge (the former being more intense than the latter).

Moreover, the right panel of Fig.~\Ref{fig:k_l1_ddcs} shows that also
RIXS spectra calculated from the BSE
at the L$_1$ and K edges overlap almost entirely, for all the incoming photon energy $\omega_1$ spanning a wide energy range of 2.6 eV across the corresponding XAS edge. 
Therefore, our detailed comparison of both XAS and RIXS spectra suggests that the less common L$_1$ edge could be equivalently utilized (if the background is not too high) at the place of the K edge. To the best of our knowledge, RIXS spectra of {\alo} have not been measured  at these edges yet. Our calculations thus represent a prediction for the direct RIXS channel.
In the following, we will analyse in detail only the RIXS spectra at the L$_1$ edge, since the K edge RIXS yields the same information.

Following Shirley and coworkers\cite{Jia_1996,RIXS_Carlisle_1999,RIXS_Shirley_2000}, we can distinguish two qualitatively different regimes for RIXS spectra (even though this separation is not sharp).
For incoming photon energies  below the XAS onset (i.e., for $\w_1 < E_{\lambda_\mu} \, \forall \lambda_\mu$),
the energy conservation  in Eq. \eqref{eq:RIXS_BSE} implies that peaks in the spectra are located at the same energy losses $\w=\w_1-\w_2=E_{\lambda_o}$ independently of the excitation energy $\w_1$.
Equivalently, the emission photon energies $\w_2 = \w_1 - E_{\lambda_o}$ increase linearly with increasing  incoming photon energies $\w_1$.
In this case, the excitations are said to have a \textit{Raman}-like behavior.
Instead, for incoming photon energies $\w_1$ above the XAS onset, the denominators in Eq. \eqref{eq:RIXS_BSE} enhance the resonant conditions $\w_1 = E_{\lambda_\mu}$ (within an energy range dictated by the inverse lifetime $\Gamma$). In this regime, the  energies of the peaks in the spectra increase linearly with the excitation energies $\omega_1$. They are thus located at constant emission energies $\w_2$, 
displaying a \textit{fluorescence}-like behavior.

The right panel Fig. \ref{fig:k_l1_ddcs} illustrates how  the BSE approach is capable to capture both regimes (the thin black lines across the different spectra indicate these two regimes).
Indeed, for excitation energies $\w_1$ below the edge,  we can identify several Raman peaks in the RIXS spectra that are located at constant energy loss, even though their shape strongly changes with $\w_1$.
Such changes in the spectrum with only small variations in $\omega_1$ cannot be explained in terms of a XAS modulation, highlighting the coherence between the absorption and emission processes.
For excitation energies $\w_1$ above the edge, instead, all the main structures of the RIXS spectra qualitatively change their behavior, displaying a linear dependence on the excitation energy, i.e., they become fluorescences.
Due to their different nature, we will analyze these two regimes in more detail, separately.  

\subsubsection{RIXS up to the XAS threshold: connection with the loss function}

\begin{figure}[th]
	\includegraphics[width=0.9\columnwidth]{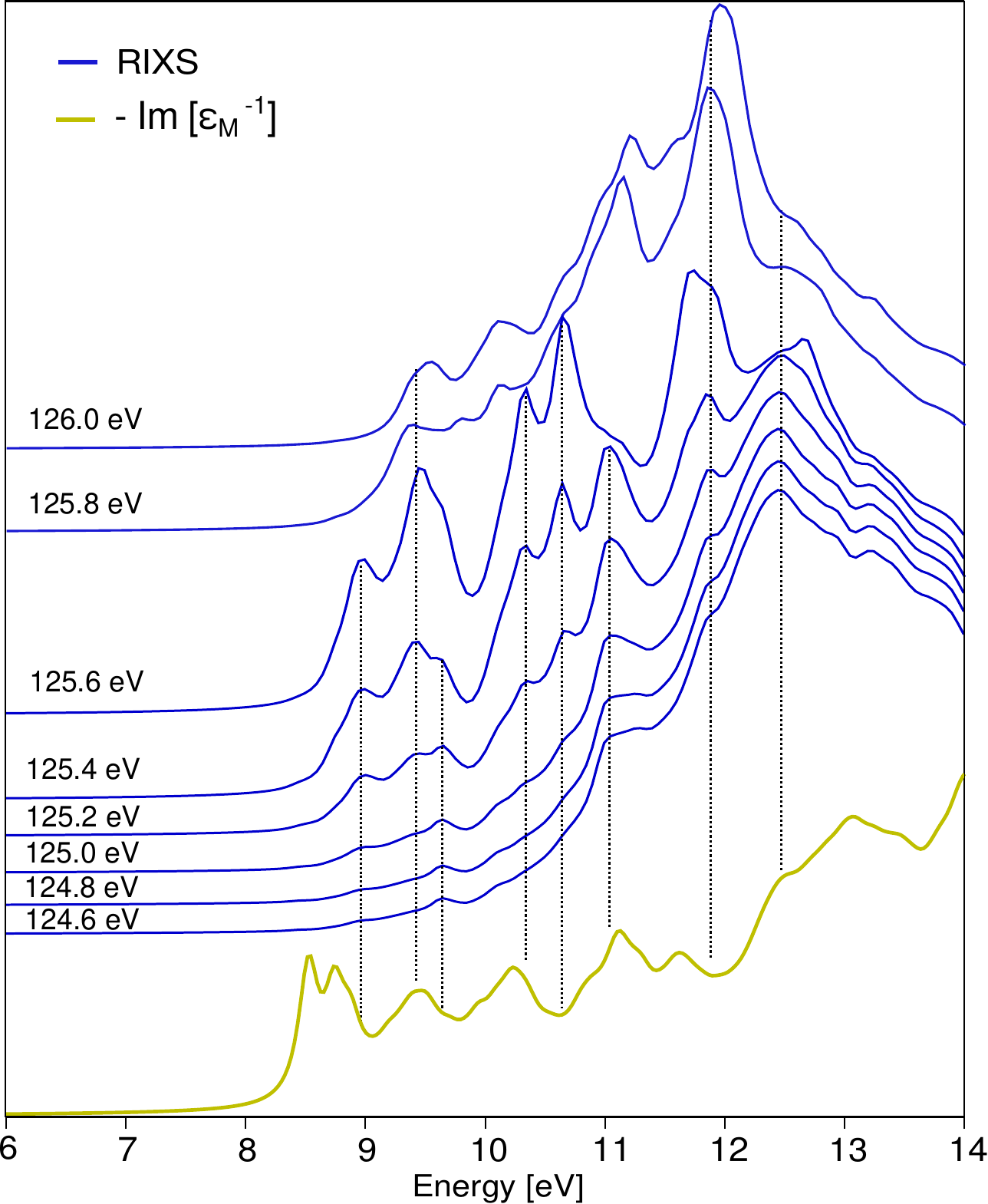}
	\caption[\columnwidth]{\label{fig:l1_rixs} BSE RIXS spectra (blue curves)  as a function of the energy loss $\omega$ at the L$_1$ edge, for excitation energies $\omega_1 < 126 $ eV. They are compared to the BSE loss function $-\textrm{Im}\epsilon_M^{-1}(\w)$ (dark yellow curve) for $\bfq\to0$ along the $x$ direction. 
	The black dotted lines connect Raman peak in RIXS spectra with corresponding inelastic losses in $-\textrm{Im}\epsilon_M^{-1}(\w)$. 
	Each spectrum has been normalized to its maximum and offset for clarity.}
\end{figure}

\begin{figure}[th]
	\includegraphics[width=1.0\columnwidth]{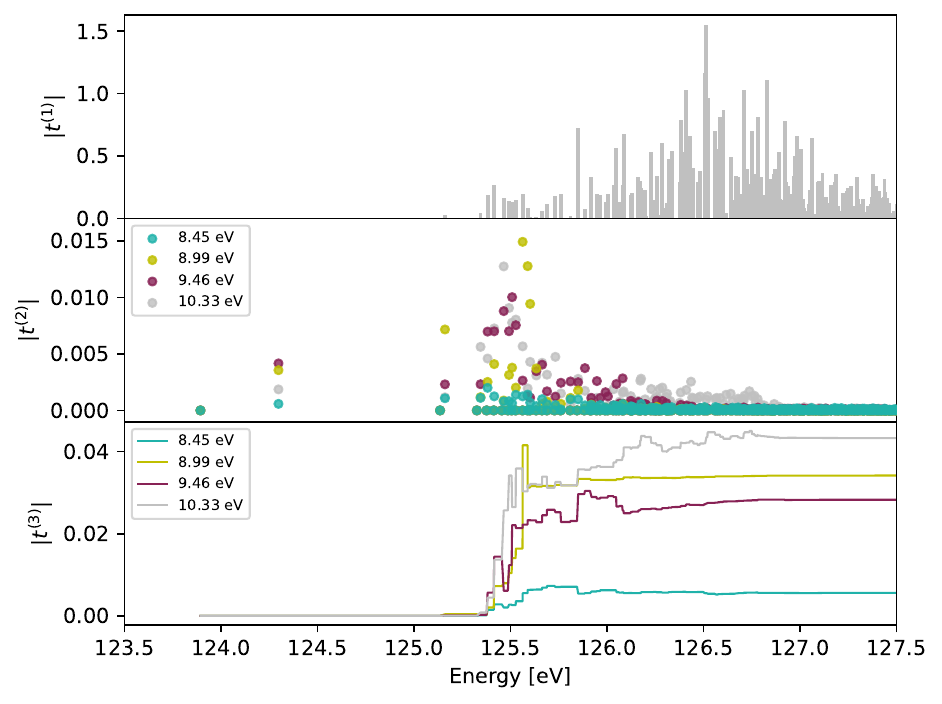}
	\caption[\columnwidth]{\label{fig:l1_t1t2t3} Absorption oscillator strengths  $|t^{(1)}_{\lambda_\mu}|$ (top) and excitation pathways $|t^{(2)}_{\lambda_o,\lambda_\mu}|$ (center) as a function of core excitation energy $E_{\lambda_\mu}$ compared to the cumulative RIXS oscillator strengths $|t^{(3)}_{\lambda_o}(E)|$ (bottom) for different exciton energies $E_{\lambda_o}$: 8.45, 8.99, 9.46 and 10.33~eV, as specified in the legends. 
	In the four cases the incoming photon energy $\omega_1$ is set to 125.6~eV.}
\end{figure}

Fig.~\ref{fig:l1_rixs} focuses on the L$_1$ RIXS spectra for incoming photon energies $\w_1$ taken every 0.2~eV between 124.6 and 126~eV, i.e., below the XAS threshold. 

The RIXS spectra are also compared to the loss function  \eqref{eq:EELS_BSE} for $\bfq\to0$.
Even though the possible excitation energies $E_{\lambda_0}$ in RIXS and in the loss function are the same (Eqs. \eqref{eq:EELS_BSE}-\eqref{eq:RIXS_BSE} share the same energy conservation terms), the corresponding peak intensities in the two cases are generally very different  (in particular, in the RIXS spectra they are also strongly dependent on the incoming photon energy $\w_1$, which enhances features at resonant energies through the denominators in Eq. \eqref{eq:RIXS_BSE}). Since moreover in a solid the possible excitation energies $E_{\lambda_0}$ form a continuum, matching the corresponding features on the basis of a direct comparison of the spectra is generally not obvious (see the vertical lines in Fig. \ref{fig:l1_rixs}).
For the {\alo} spectra in Fig. \ref{fig:l1_rixs}, in particular, the most evident discrepancy are the two peaks in the loss function  below 9~eV that are missing in the RIXS spectra. 
They correspond to two bound excitons\cite{L1_XAS_Urquiza} located inside the band gap of {\alo}. We can therefore wonder: Why aren't they visible also in the RIXS spectra? And, more in general, how can we use the information from the BSE eigenvalues and eigenfunctions to analyse RIXS spectra and connect them with the loss function?

In order to address these questions in detail and analyze the interference between core and valence excitons, we introduce the cumulative oscillator strength $t^{(3)}$ for the peak of energy $E_{\lambda_0}$ in the RIXS spectrum:
\begin{equation} 
	t^{(3)}_{\lambda_o}(E) =  \sum_{\lambda_\mu}^{E_{\lambda_\mu}<E} \frac{t^{(1)}_{\lambda_\mu} \ t^{(2)}_{\lambda_o,\lambda_\mu}}{\omega_{1} - E_{\lambda_\mu} + i\Gamma/2}.
\label{eq:t3}
\end{equation}

For $E\to\infty$, $|t^{(3)}_{\lambda_o}|$ yields the peak intensity in the RIXS spectrum (calculated for incoming photon energy $\w_1$) associated to the excitation energy $E_{\lambda_0}$.
Its plot as a function of the energy $E$ reflects the interference between absorption and emission processes, represented by the terms $t^{(1)}$ and $t^{(2)}$ at the numerator of Eq. \eqref{eq:t3}, respectively. 
Since they are complex numbers, the interference can be constructive or destructive, displaying the many-body character of the excitations\cite{L1_XAS_Urquiza,Lorin2021}.
At the same time also the denominator of Eq. \eqref{eq:t3} may cross  zero for $E_{\lambda_\mu}\sim\w_1$, expressing the resonant nature of RIXS.
As a result, $|t^{(3)}_{\lambda_o}|$ is generally not a monotonic function of the energy $E$. 

The top and middle  panels of Fig.~\ref{fig:l1_t1t2t3} respectively show the absorption oscillator strenghts $|t^{(1)}_{\lambda_\mu}|$ (see Eq. ~\eqref{eq:t1}), and the excitation pathways $|t^{(2)}_{\lambda_o,\lambda_\mu}|$ for few selected valence excitons $E_{\lambda_o}$  (see Eq. ~\eqref{eq:t2}), as a function of the $E_{\lambda_\mu}$ core excitation energy.
To understand why the first peak in the loss function is not visible in RIXS, in the bottom panel of Fig.~\ref{fig:l1_t1t2t3} we analyze the cumulative oscillator strenght $|t^{(3)}_{\lambda_o}(E)|$. 
In this analysis, the incoming photon energy $\omega_1$ is set to 125.6~eV, corresponding to the higher energy for which Raman features are noticeable.
The behavior of the valence exciton  with energy $E_{\lambda_{o}} = 8.45$~eV is compared  with those of the excitons at energies $E_{\lambda_{o}} =$ 8.99~eV, 9.45~eV, and 10.33~eV, corresponding to the first three visible peaks in the RIXS spectra.
For all the valence excitons $E_{\lambda_o}$,  the excitation pathways  $|t^{(2)}_{\lambda_o,\lambda_\mu}|$ are  not zero at low excitation energies $E_{\lambda_{\mu}} < 125$ eV, where the absorption oscillators  $|t^{(1)}_{\lambda_\mu}|$ are instead negligible, due to the forbidden nature of the transition between the Al $2s$ core level and the bottom of the conduction band, primarily of Al $3p$ character. Accordingly, they do not contribute to the RIXS spectrum. 
For the exciton $E_{\lambda_{o}} = 8.45$~eV, the values of the other excitation pathways $|t^{(2)}_{\lambda_o,\lambda_\mu}|$ are very small, canceling out the contributions of the absorption oscillators $|t^{(1)}_{\lambda_\mu}|$ for all other $E_{\lambda_{\mu}}$ energies. 
The resulting oscillator strength $|t^{(3)}_{\lambda_o}(E)|$ is therefore always very small, and as a result the exciton is not visible in the RIXS spectrum. 
For the other valence excitons $E_{\lambda_{o}}$ corresponding to the visible peaks in the RIXS spectrum, instead, the emission  contributions $|t^{(2)}_{\lambda_o,\lambda_\mu}|$ provide constructive interference with the absorption oscillators $|t^{(1)}_{\lambda_\mu}|$ at higher $E_{\lambda_{\mu}}$ energies. 
As a result, the peaks are visible in the RIXS spectrum.

From this analysis, we can therefore conclude that the lowest-energy peaks in the loss function (see the yellow curve in Fig. \ref{fig:l1_rixs}) are not visible in the RIXS spectra because the first absorption step, corresponding to the excitation from core $2s$ level to the bottom conduction states, is forbidden. 
In the Al L$_1$ and K edges of {\alo}, RIXS is not able to probe the lowest energy valence exciton states involving combinations of top-valence and bottom-conduction states, but only excitons with higher energies ($\gtrsim$ 9 eV) stemming from higher conduction states.

\subsubsection{RIXS at and beyond the XAS threshold: excitonic effects fading away and connection with XES}

\begin{figure}
	\includegraphics[width=0.9\columnwidth]{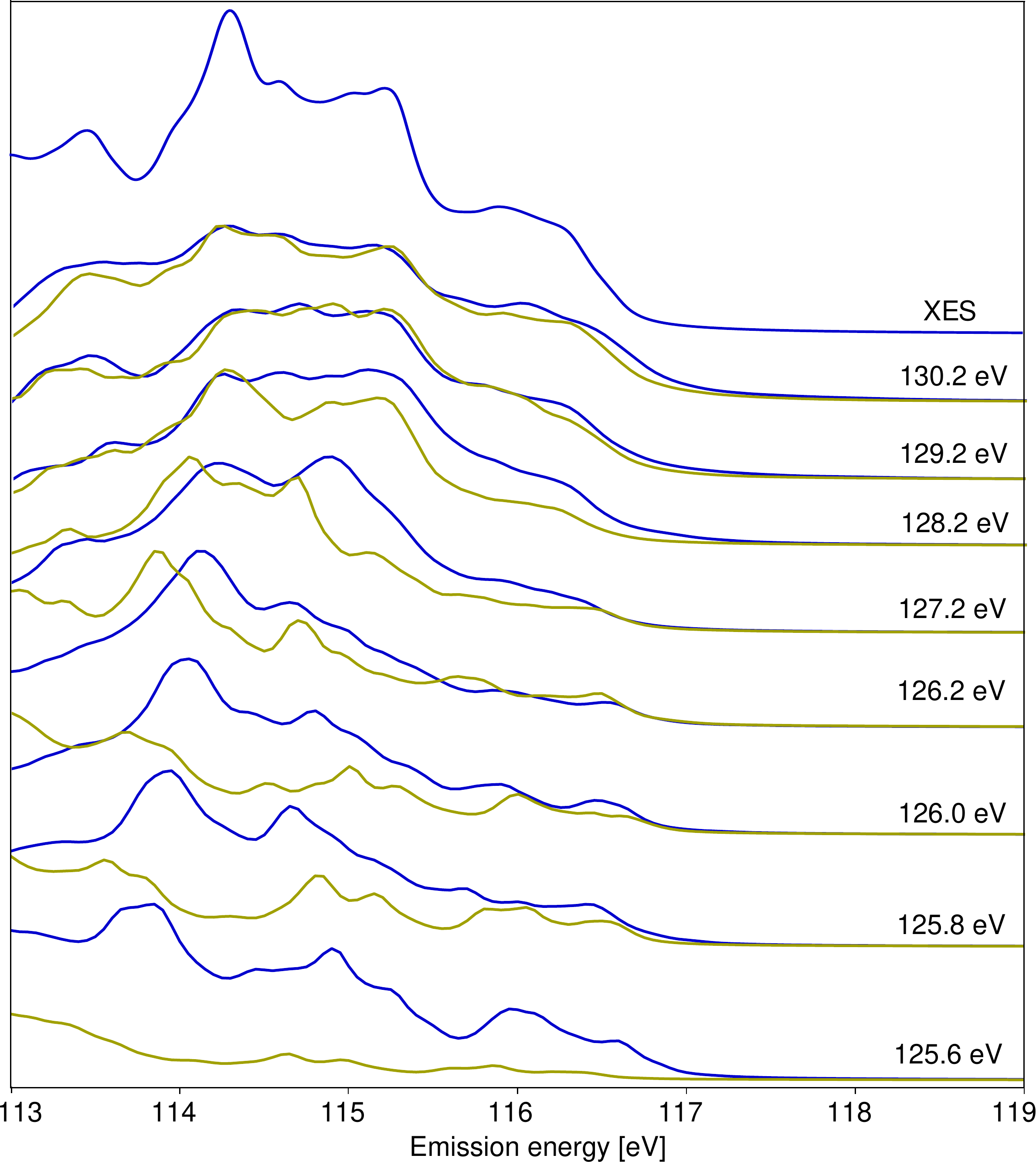}
	\caption{\label{fig:l1_rixs_xes} Comparison between BSE (blue curves) and IPA (dark yellow curve) RIXS spectra as a function of the emission energy $\w_2$, for different excitation energies $\w_1$ across the XAS Al L$_1$ threshold and above it. The fluorescence features are also compared with the XES spectrum (blue curve at the top) at the same Al edge. The RIXS spectrum has been calculated for $\evec_1$ and $\evec_2$ along the $x$ direction, and the XES spectrum for $\qvec\to0$ in the $x$ direction.  
	In the plot, each spectrum has been normalized to its maximum and offset for clarity. }
\end{figure}
\begin{figure}
	\includegraphics[width=0.9\columnwidth]{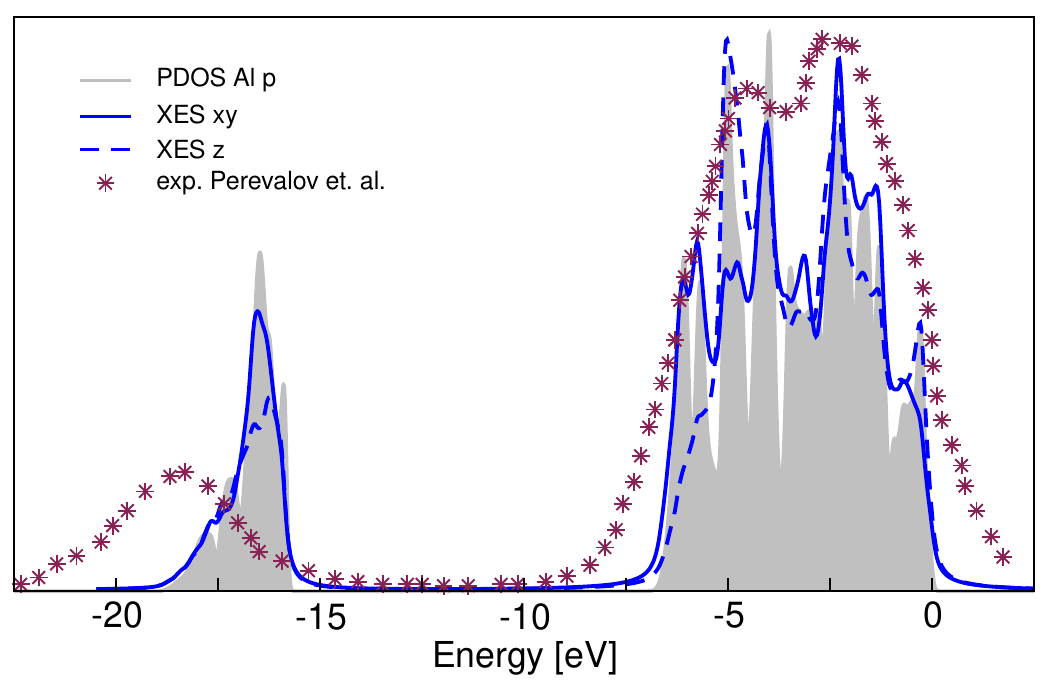}
	\caption{\label{fig:pdos_xes} Comparison between experimental x-ray emission spectrum at Al K edge extracted from Ref.\cite{K_XES_Perevalov} and calculated XES within IPA for the two polarization directions $xy$ and $z$. 
	The calculated XES can be well described by the Al $3p$ component of the PDOS, as highlighted in the figure. 
	In all cases, the top of the valence band in the PDOS and the XES have been set to 0 eV.}
\end{figure}

Fig.~\ref{fig:l1_rixs_xes} presents BSE and IPA RIXS spectra for excitation energies $\w_1$ across and above the XAS Al L$_1$ threshold. The two series of spectra have been obtained from Eqs. \eqref{eq:RIXS_BSE} and \eqref{eq:RIXS_IPA}, respectively.
Note that, in order to highlight fluorescence features, here the spectra have been plotted as a function of the emission photon energy $\w_2$, instead of the energy loss $\w$, as shown in Figs.~\ref{fig:k_l1_ddcs}-\ref{fig:l1_rixs}.

Below the threshold and for energies up to 128.2~eV large differences between IPA and BSE spectra are noticeable.
This implies that strong excitonic effects impact RIXS in {\alo}, making it impossible an interpretation based only on a band-structure picture.

One should therefore adopt a many-body picture, where the intermediate and final states correspond to a superposition of many single-particle states at different crystal momenta and from different bands. 

On the contrary, for incoming photon energies $\w_1$ well above the XAS threshold, the IPA and the BSE provide quite similar results. In this regime, the features observed at constant emission energy are indicative of a two-step fluorescence process, where absorption and emission are independent.  
Consequently, the absorption process (represented by the $t^{(1)}$ contributions in Eq.~\eqref{eq:RIXS_BSE}) acts as a scaling factor and the RIXS spectrum is characterised by fluorescence features (determined by the $t^{(2)}$ contributions in Eq.~\eqref{eq:RIXS_BSE}). 
At high incoming photon energies, since excitonic effects are negligible, the BSE RIXS~\eqref{eq:RIXS_BSE} becomes similar to IPA RIXS~\eqref{eq:RIXS_IPA}, and  since absorption and emission are decoupled, the RIXS spectrum also becomes very close to the XES spectrum ~\eqref{eq:XES_IPA}.

Moreover, as evidenced by Fig.~\ref{fig:pdos_xes}, the  XES spectrum can be very well described by the angular $p$ component of Al in the PDOS for the valence band. 
The PDOS shows peaks between -20~eV and -15~eV corresponding to aluminum $3p$ states hybridized with the oxygen $2s$ states. 
Separated by a large gap of $\sim$8~eV, starting at $\sim$-7~eV up to the maximum of the valence band set at 0 eV, one finds Al $3p$ states mostly hybridized with O $2p$ states.
Even though the separation between the O $2s$ and $2p$ groups of bands is underestimated in the LDA, the calculations reproduce very well the experimental XES spectrum from Ref.\cite{K_XES_Perevalov}, including the two-peak structure in the O $2p$ valence band.

Even in a material like {\alo} where strong excitonic effects are at play (for both core and valence excitations), we can therefore conclude that at high photon energy $\w_1$ RIXS is still capable to provide information about band-structure properties.

\begin{figure*}
	\includegraphics[width=1.7\columnwidth]{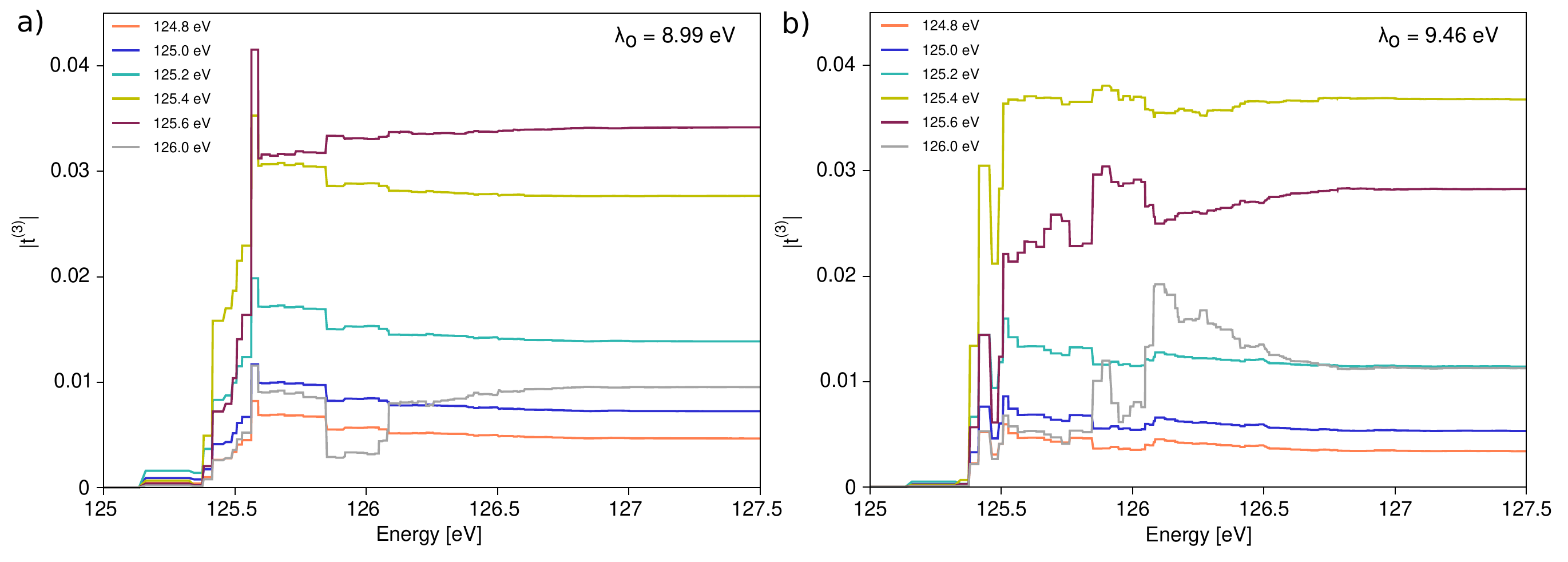}
	\caption[\columnwidth]{\label{fig:l1_t3_raman} Cumulative function $|t^{(3)}_{\lambda_o}(E)|$ for different incoming photon energies within the Raman regime. 
	The two panels represent different optical exciton energies $E_{\lambda_o}$, associated to RIXS peaks at energy loss of (a)~8.99~eV and (b)~9.46~eV, as seen in Fig.~\ref{fig:l1_rixs}.}
	\includegraphics[width=1.7\columnwidth]{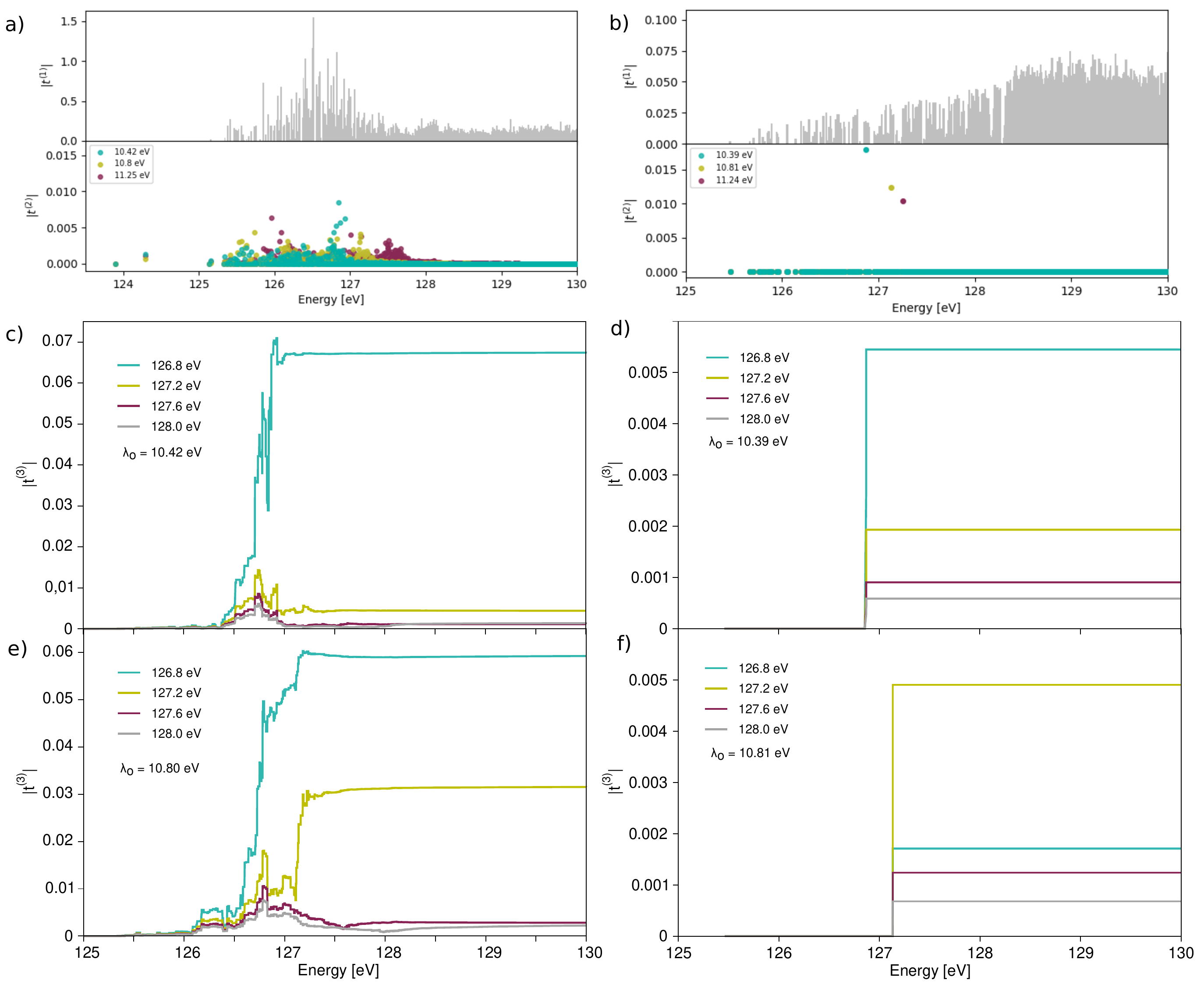}
	\caption[\columnwidth]{\label{fig:l1_t3_fluorescence} Absolute value of the absorption oscillator strength $|t^{(1)}_{\lambda_{\mu}}|$ and the excitation pathways $|t^{(2)}_{\lambda_{o},\lambda_{\mu}}|$ at different $E_{\lambda_{o}}$ as specified in the legends for (a)~BSE and (b)~IPA RIXS. Cumulative function $|t^{(3)}_{\lambda_o}(E)|$ for different incoming photon energies in the fluorescence regime obtained with the BSE for optical excitation energies $E_{\lambda_{o}}$ of (c)~10.42~eV and (e)~10.80~eV. Similarly, in the right panels it has been plotted the $|t^{(3)}_{\lambda_o}(E)|$ calculated within IPA for $E_{\lambda_{o}}$ equal to (d)~10.39~eV and (f)~10.81~eV.}
\end{figure*}

In order to understand how Raman features in RIXS spectra give way to fluorescence features as the incoming photon energy $\w_1$ increases, we examine the dependence of the oscillator strenghts $t^{(3)}_{\lambda_o}$  \eqref{eq:t3} on $\w_1$, for specific optical excitons $E_{\lambda_{o}}=$~8.99~eV and 9.46~eV, as shown in Fig.~\ref{fig:l1_t3_raman}.
The two plots show that for incoming photon energies below the XAS threshold (125.6~eV), the interference patterns are relatively constant. On the other hand, the intensity increases accordingly with the resonance between $\omega_1$  and the core exciton energies $E_{\lambda_{\mu}}$, reaching the highest value at $\omega_1=$~125.4--126.6~eV.
Beyond this energy, the oscillator strength decays as a result of two effects: the loss of resonance (for the two specific optical exciton $\lambda_{o}$ considered here) and the change in the interference pattern produced by the sign in the denominator of Eq.~\eqref{eq:RIXS_BSE}. 
As a result, $|t^{(3)}_{\lambda_{o}}(E)|$ becomes negligible in the case of the first peak, at an energy loss of 8.99~eV, while it remains visible but with lower intensity at 9.46~eV. 
This explains the RIXS features of Fig.~\ref{fig:l1_rixs}, where the first peak disappears after $\omega_1=$125.6~eV and the second peak is visible up to $\omega_1=$126.0~eV. 

The second observation that can be extracted from the plots is that even though the absorption oscillator strength $t^{(1)}_{\lambda_{\mu}}$ and excitation pathways $t^{(2)}_{\lambda_{o},\lambda_{\mu}}$ are the same for a given $\lambda_{o}$, when the incoming photon energies are beyond the XAS threshold, the contributions to $|t^{(3)}_{\lambda_o}|$ tend to shift to higher energies as $\omega_1$ increases. 
This trend is more noticeable in the fluorescence regime, which will be explained below. 

Fig~\ref{fig:l1_t3_fluorescence} shows the absolute value of the absorption oscillator strengths $|t^{(1)}_{\lambda_{\mu}}|$ and the excitation pathways $|t^{(2)}_{\lambda_{o},\lambda_{\mu}}|$ for (a)~BSE and (b)~IPA RIXS, at three different energy losses: 10.4, 10.8 and 11.2~eV.  
The resulting cumulative oscillator strengths $|t^{(3)}_{\lambda_o}|$ calculated for BSE RIXS at two $E_{\lambda_{o}}$ values are displayed in panels (c)~(10.42~eV) and (e)~(10.80~eV). 
Similarly, the $|t^{(3)}_{\lambda_o}|$ calculated within IPA are displayed in the panels (d) for $\lambda_{o}=$~10.39~eV and (f) for $\lambda_{o}=$~10.81~eV.
The figure highlights the fact that, given a specific optical exciton $\lambda_{o}$, the differences in the $|t^{(3)}_{\lambda_o}|$ with incoming photon energy $\omega_1$ are only given by the effect of the denominator of Eq.~\eqref{eq:RIXS_BSE}, as discussed above. 

Here it is interesting to notice that the largest contributions of the excitation pathways $|t^{(2)}_{\lambda_o,\lambda_\mu}|$ (panels (a) and (b)) tend to be centered  around higher energies $E_{\lambda_{\mu}}$ as the energy of the valence excitons $E_{\lambda_{o}}$ increases (this has been also observed in Fig.~\ref{fig:l1_t1t2t3}). 
These excitons stem from electron-hole transitions of higher energies, originating from valence and conduction states farther away from the top valence and bottom conduction, respectively. 
Correspondingly, a larger excitation energy  $E_{\lambda_{\mu}}$ is required to probe those higher-energy conduction states.

If one compares the RIXS oscillator strengths for the BSE (panels (c) and (e)) and IPA (panels (d) and (f)), it is possible to conclude that the blueshifts in the energy  contributions to $|t^{(3)}_{\lambda_o}|$ for increasing $\omega_1$ are the result of the excitonic effects, which give more weight to higher core excitons as $\omega_1$ increases, while in the IPA picture the jump in the $|t^{(3)}_{\lambda_o}|$ is given at a constant energy $\lambda_{\mu}$ for all $\omega_1$.

If we compare the (c)-(f) panels in Fig~\ref{fig:l1_t3_fluorescence}, it is noticeable how the RIXS oscillator strength $|t^{(3)}_{\lambda_o}|$ for high $\omega_1$ leads to negligible values at low energy-loss peaks $\lambda_{o}$ (panels (c) and (d)), while they become more prominent as the energy loss increases (panels (e) and (f)). 
This analysis provides the last piece to understand the movement of peaks in the fluorescence regime.
The fact that the excitation pathways $|t^{(2)}_{\lambda_{o},\lambda_{\mu}}|$ (panels (a) and (b)) shift to higher core energies $\lambda_{\mu}$ as one moves to higher optical excitation energies $\lambda_{o}$ (or energy loss) imposes resonance (according to Eq~\eqref{eq:RIXS_BSE}) at higher incoming photon energy $\omega_1$. Consequently, the increase of  $\omega_1$ causes that RIXS oscillator strength $|t^{(3)}_{\lambda_{o}}|$ becomes more intense at higher $E_{\lambda_{o}}$.

As a summary, one can highlight that within the many body picture, $\omega_1$ has two main effects: 1) before the XAS edge, it acts as a scaling factor, leaving the role of the interference to the interplay between $t^{(1)}$ and $t^{(2)}$ at each optical exciton energy $E_{\lambda_{o}}$; and 2) for incoming photon energies in resonance and beyond the XAS edge, it determines for each core exciton $\lambda_{\mu}$, whether the interference between $t^{(1)}_{\lambda_{\mu}}$ and $t^{(2)}_{\lambda_{o},\lambda_{\mu}}$ will be constructive or destructive according to the sign.

\section{\label{sec:conclusions} Conclusions}

We have presented an in-depth analysis of RIXS spectra, considering coherence and excitonic effects throughout the process. Additionally, we have established several connections between RIXS and complementary spectroscopy techniques that assess neutral excitations in materials, specifically XAS, XRS, and XES. We have applied the BSE approach to study core and semicore excitations of corundum \alo, a widely used material due to its optical and structural properties.

The comparison between experimental Al K  edges XAS spectra and calculated spectra at different polarization directions has demonstrated that excitonic effects are crucial to reproduce most features of the experimental data. The experimental prepeak (A) observed at $\sim$1566 eV is a dark exciton representing dipole-forbidden transitions in the calculations. 
However, it becomes a bright exciton in the calculated loss functions at finite momentum transfers, which includes multipole contributions.  

Remarkably, our study reveals that both XAS and RIXS spectra, even when calculated within the BSE, at the K and L$_1$ edges exhibit a high degree of agreement. This result suggests the use of soft X-ray techniques for studying the L$_1$ edge and extracting the same information as the traditionally explored K edge. 

For excitation energies below the XAS onset, the RIXS spectra show a Raman-like behavior, where the energy losses remain constant as the incident photon energy $\omega_1$ is varied. 
The comparative analysis of the Raman losses between RIXS and the loss function sheds light on differences in selection rules and intensity enhancements between the two techniques. 
Beyond the XAS threshold, RIXS displays a two-step fluorescence behavior, with peaks shifting in accordance with $\omega_1$, highlighting the loss of coherence between the absorption and the emission processes. 
The agreement with XES emphasizes the similarities between fluorescence and emission.
In the same fashion, excitonic effects become weaker for excitation energies
$\omega_1$ well above the threshold.
Therefore, if one wishes to probe valence excitons  by RIXS, excitation energies $\w_1$ below or at the edge should be selected.
Conversely, if band-structure properties are the target, higher excitation energies $\w_1$ should be preferred.

Altogether, these findings contribute to a deeper understanding of the RIXS process, its behavior in different energy regimes, and its connection with other spectroscopic techniques, facilitating further insights into materials' electronic properties.

\begin{acknowledgments}
We acknowledge valuable discussions with Christian Vorwerk.
We thank the French Agence Nationale de la Recherche (ANR) for financial support (Grant Agreements No.  ANR-19-CE30-0011). Computational time was granted by GENCI (Project  No.  544).   
\end{acknowledgments}

\bibliography{bib}

\end{document}